# ENERGY DISPERSION COMPENSATION AND BEAM LOADING IN X-BAND LINACS FOR THE JLC/NLC[§]

R.M. Jones, V.A. Dolgashev, R.H. Miller, C. Adolphsen and J.W. Wang

Stanford Linear Accelerator Center, 2575 Sand Hill Road, Menlo Park, CA, 94025

## Abstract

The shape of an RF pulse is distorted upon propagating through an X-band accelerator structure due to dispersive effects. This distortion together with beam loading introduce energy spread between 192 bunches. In order to minimize this energy spread we modify the input RF pulse shape. The pulse propagation, energy gain, and beam loading are modelled with a mode-matching computer code and a circuit model. A 2D model and a circuit model of a complete 60 cm structure, consisting of 55 cells and input and output couplers is analyzed. This structure operates with a $5\pi/6$ phase advance per cell. Dispersive effects for this structure are more significant than for previously studied $2\pi/3$ phase advance accelerating structures. Experimental results are compared with the theoretical model and excellent agreement is obtained for the propagation of an RF pulse through the structure.

[§] Supported by Department of Energy grant number DE-FG03-93ER40759

# ENERGY DISPERSION COMPENSATION AND BEAM LOADING IN X-BAND LINACS FOR THE JLC/NLC [1]

R.M. Jones, V.A. Dolgashev, R.H. Miller, C. Adolphsen and J.W. Wang; SLAC, USA

Abstract

The shape of an RF pulse is distorted upon propagating through an X-band accelerator structure due to dispersive effects. This distortion together with beam loading introduce energy spread between 192 bunches. In order to minimize this energy spread we modify the input RF pulse shape. The pulse propagation, energy gain, and beam loading are modelled with a mode-matching computer code and a circuit model. A 2D model and a circuit model of a complete 60 cm structure, consisting of 55 cells and input and output couplers is analyzed. This structure operates with a $5\pi/6$ phase advance per cell. Dispersive effects for this structure are more significant than for previously studied $2\pi/3$ phase advance accelerating structures. Experimental results are compared with the theoretical model and excellent agreement is obtained for the propagation of an RF pulse through the structure.

## 1. INTRODUCTION

The NLC/JLC (Next Linear Collider/Japanese Linear Collider) baseline design now consists of eight 60cm accelerating structures attached to a girder (of which there are several thousand). Each cell in the structure has a phase advance of $5\pi/6$. Previously $2\pi/3$ was the phase advance of choice, but a higher phase advance was chosen in order to reduce the group velocity of the fundamental mode wave, which was thought to reduce any electrical breakdown occurring in the structures [1]. A CAD view of several cells in H60VG3 is shown in fig. 1. This accelerating structure consists of 55 cells and has been carefully designed to reduce the electromagnetic fields on the surface of the copper, to minimize the pulse temperature heating effects (by shaping the cavity-manifold slots into a "pie" shape).

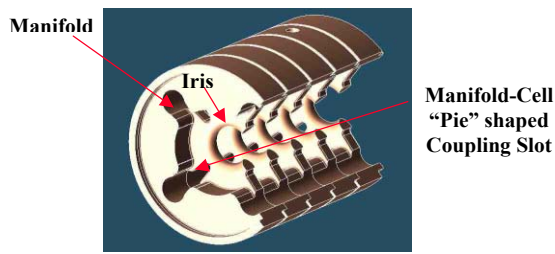

Figure 1: Cross-sectional view of a CAD drawing of the H60VG3 accelerating structure

Sending an rf pulse down the accelerator resulted in a noticeable distortion in the shape of the pulse output from 55 cells and this is discussed in section 2. Section 3 delineates the effect of beam loading on the overall pulse shape and means to overcome the energy deviation over the pulse train caused by this dispersive effect.

## 2. DISPERSION IN TW STUCTURES

Shown in fig 2 is the result of an experiment to measure the pulse shape after transiting through all 55 cells of the TW (Travelling Wave) structure H60VG3. Distortion is evident at both the trailing and leading edge of the pulse.

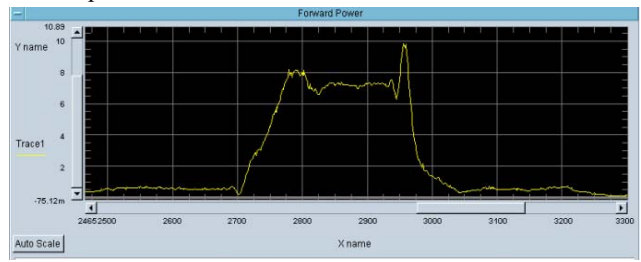

Figure 2: Oscilloscope measurement of the output pulse for a 240ns input rectangular pulse with a 10ns rise and fall time injected into H60VG3

To simulate the propagation of the pulse we utilize the 0 and $\pi$ frequencies from 5 cells in a 3D simulation of cells with boundary conditions corresponding to an infinitely periodic structure. From these points we obtain the corresponding dispersion curves for these cells and additional cells are obtained by interpolation. We use a circuit model [2] to simulate the propagation of a pulse through this structure and also through constant impedance structures.

Prior to obtaining the pulse shape the accelerator must matched to the attached couplers and this is simulated by varying the end cell frequencies and Q values. Fig. 3 illustrates the frequency behaviour of the reflection coefficients after the structure has been matched. In all cases to achieve this reflection coefficient we utilize [3] and apply it to the circuit model to obtain minimize the internal reflection coefficient of the accelerator. The pulse shape after propagating through all 55 cells is shown in fig 4 for three phase advances: $2\pi/3$, $3\pi/4$ and $5\pi/6$. Before electrical breakdown and the consequent damage to the structures was discovered [4], all DDS (Damped and Detuned Structures) were fabricated with a $2\pi/3$ phase advance per cell. Little pulse distortion is seen for this case. The reduced group velocity structure that was measured experimentally (fig 2.) corresponds to the simulation in fig 4 with a 5π/6 phase advance per cell. The distortion recorded in the experiment is well represented by the simulation and it is clear that the pulse distortion occurs due to dispersion in the structure. The intermediate phase advance of 3π/4 gives rise to very little distortion in the pulse shape and it is shown because it corresponds to another potential structure that has been under consideration for the NLC. In all cases shown in figs. 3 and 4 a circuit model has been used. However, a mode matching method [5] applied to the 5π/6 case shows very similar pulse distortion. We utilize this mode matching method to calculate energy gained by 192 electron bunches traversing the structure and this is described in the following section

[1] Work supported by the U.S. DOE Contract No. DE-AC03-76SF00515.

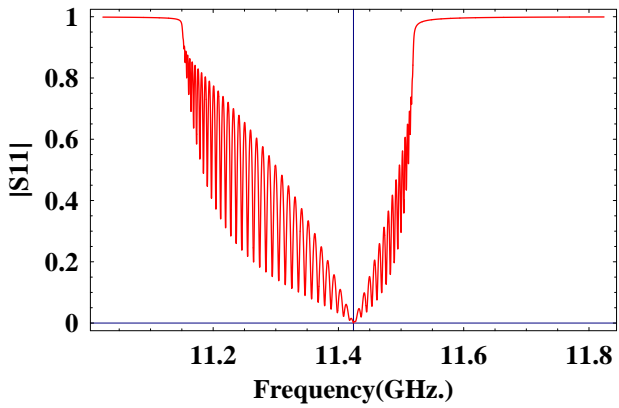
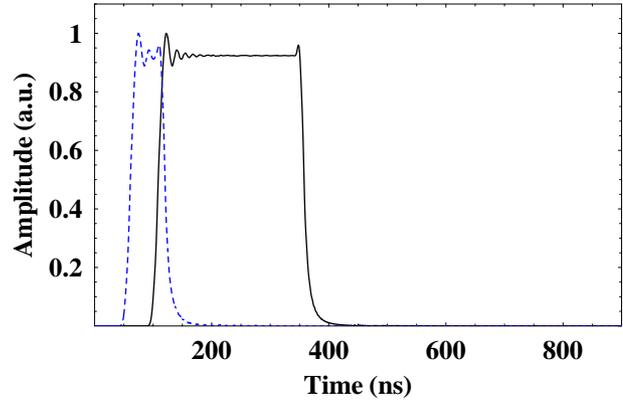
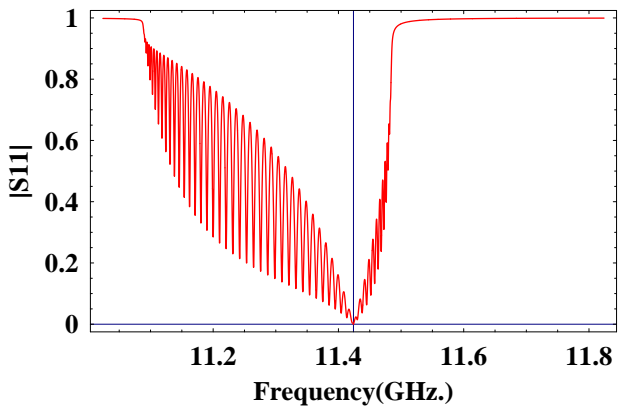
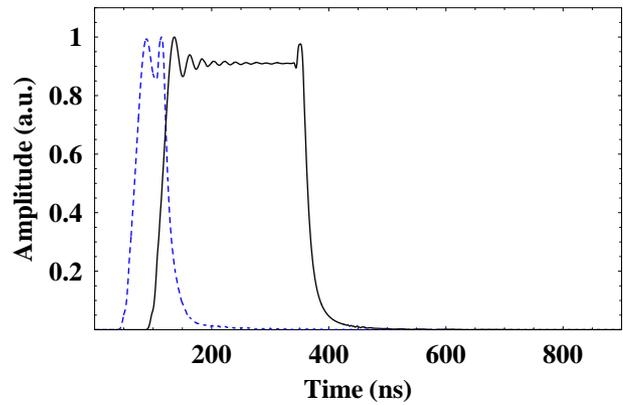
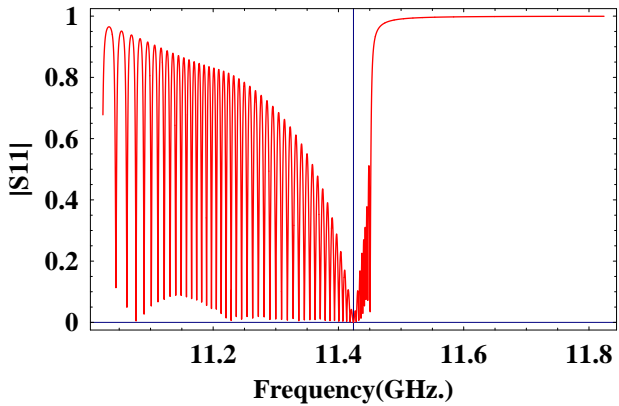
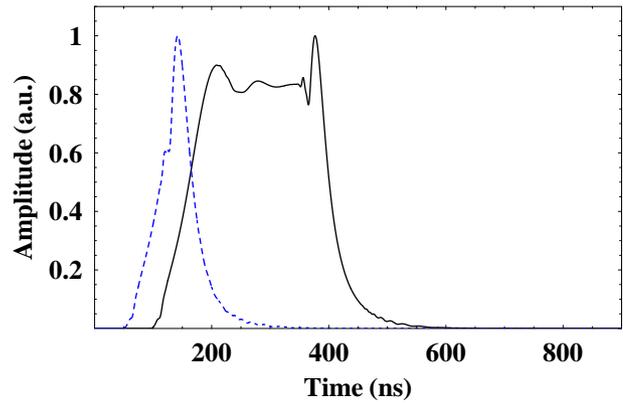

Figure 3: Reflection coefficient of the input to several travelling wave accelerator structures each of which consists of 55 cells. The uppermost consists of a constant impedance structure with a phase advance per cell of 120 degrees, the middle to a constant impedance with 135 degrees phase advance and, the lowermost to H60VG3 in which the phase advance is 150 degrees and the cell parameters vary smoothly along the accelerator structure.

Figure 4: A rectangular pulse with a fall and rise time of 10ns and a flat-top of 240ns and 50ns is propagated through the accelerating structures described in fig 3. The final output pulse for each structure is illustrated above (blue and dashed for the shorter input pulse).

## 3. BEAM LOADING AND ENERGY DISPERSION COMPENSATION

Multiple bunches of electrons passing through an accelerating structure load down the accelerating field and consequently the field is reduced. In order to reduce this droop in the field the initial RF pulse is ramped up as the structure is filled with electromagnetic field. Here we calculate the effect of beam loading from the impedance of the beam [5] and the beam loading, together with the total pulse are shown in fig 5. Over all bunches the energy deviation from bunch-to–bunch is quite small. However, the energy deviation at the final focus of the collider is required to be very small, namely 0.25% RMS [6]. For this reason we show the energy deviation along the pulse train on an expanded scale in fig 6. The set of points labelled 0 corresponds to this energy deviation and the maximum deviation is approximately 0.75%.

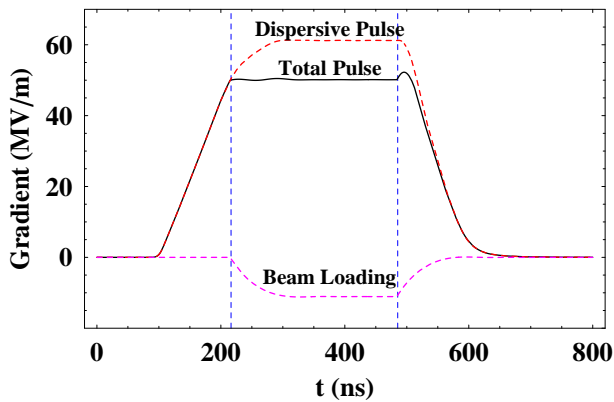

Figure 5: Individual components of a beam-loaded pulse that reaches cell 55 of H60VG3. The dashed vertical (blue) lines indicate the region in which all bunches are accelerated.

To compensate for this deviation in the energy over the bunch train we investigated reshaping the input RF pulse to the structure. A fraction of the derivative of the energy deviation over train of bunches is added to the original input pulse and this fraction is varied until the energy deviation is minimized. The set of points (in blue) labelled with "1" in fig 6 corresponds to the new energy deviation along the bunch train; the maximum deviation is below 0.2%. Application of the same iterative method again gives the set of points "2" (magenta) and this reduces the energy deviation still further, to less than 0.1%. Further application of this technique allows the energy deviation to be reduced to less than 0.05%. The input pulse prescribed from this process is shown in fig 7.

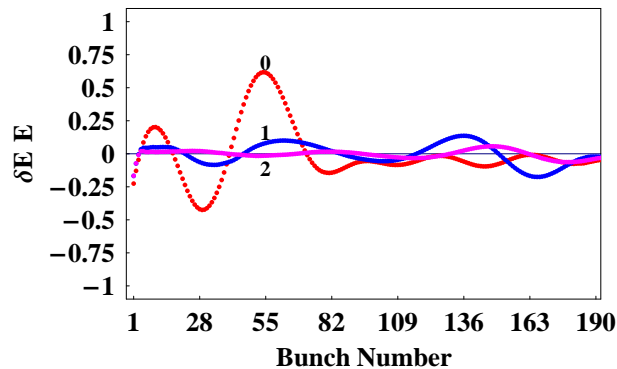

Figure 6: Percentage energy deviation along the train of electron bunches for each stage of the iteration

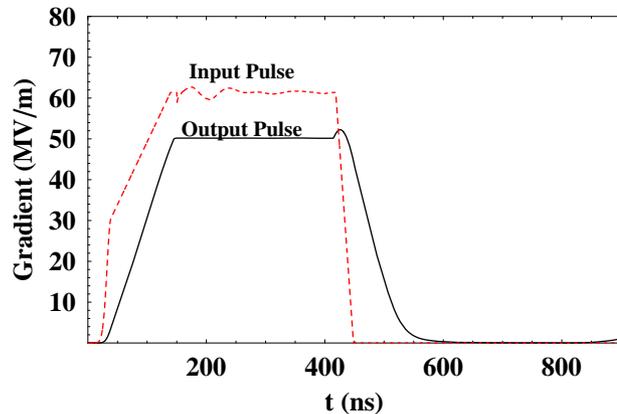

Figure 7: Input pulse to compensate for energy deviation over the pulse train together with final output pulse

Manipulating the phase of the klystrons that feed the structure will allow this pulse shape to be obtained [7].

This energy compensation method results in a small loss of energy as the pulse is compensated over the course of the bunch train. We are also investigating applying a similar compensation scheme over the initial ramp of the pulse and this will minimize the loss in overall efficiency.